\documentclass[]{aastex702}

\usepackage{xcolor}

\shorttitle{A Salpeter-like Filament Linear Density Function}
\shortauthors{Zhang, Men'shchikov \& Li}

\begin{document}

\title{A Salpeter-like Filament Linear Density Function across Nearby Molecular Clouds}

\author[orcid=0000-0002-7254-669X,gname=Guo-Yin,sname=Zhang]{Guo-Yin Zhang}
\altaffiliation{These authors contributed equally to this work.}
\affiliation{National Astronomical Observatories, Chinese Academy of Sciences, Beijing 100101, China}
\email[show]{zgyin@nao.cas.cn}
\correspondingauthor{Guo-Yin Zhang, Alexander Men'shchikov}

\author[orcid=0000-0001-5201-5595,gname=Alexander,sname=Men'shchikov]{Alexander Men'shchikov}
\altaffiliation{These authors contributed equally to this work.}
\affiliation{Universit\'{e} Paris-Saclay, Universit\'{e} Paris Cit\'{e}, CEA, CNRS, AIM, 91191 Gif-sur-Yvette, France}
\email[show]{alexander.menshchikov@cea.fr}

\author[orcid=0000-0002-9331-8194,gname=Jin-Zeng,sname=Li]{Jin-Zeng Li}
\affiliation{National Astronomical Observatories, Chinese Academy of Sciences, Beijing 100101, China}
\email[hide]{ljz@nao.cas.cn}

\begin{abstract}
The high-mass slope of the stellar initial mass function (IMF), first measured by Salpeter in 1955, appears nearly universal across star-forming environments, yet its physical origin remains unknown. Using the multiscale extraction method \textit{getsf}, we measure the filament linear density function (FLDF) across seven nearby molecular clouds (140--920\,pc) spanning a wide range of star-forming activity. The FLDF slope shows no systematic trend with spatial scale. Combining all clouds and scales, the composite FLDF follows a power law ${\rm d}N/{\rm d}\log\Lambda \propto \Lambda^{-\alpha}$ with $\alpha \approx 1.40$, close to the Salpeter value of $1.35$, and the same composite slope is found in every cloud despite an order-of-magnitude spread in their supercritical-filament fractions. 
The close similarity between the composite FLDF slope and the Salpeter value suggests a possible physical link between the hierarchical filamentary structure of the cold interstellar medium and the IMF.
\end{abstract}


\section{Introduction}
Why the stellar initial mass function (IMF) -- the distribution of stellar masses at birth -- follows a power law ${\rm d}N/{\rm d}\log M \propto M^{-1.35}$ at high masses, first identified by \citet{Salpeter1955} and remarkably universal across diverse environments \citep{Bastian+2010, Kroupa+2026}, remains a defining problem in astrophysics \citep{Kroupa2002, Chabrier2003}. \textit{Herschel} surveys established that dense filaments dominate the mass budget of molecular clouds and host most prestellar cores and star formation \citep{Andre+2010, Menshchikov+2010, Arzoumanian+2011, Konyves+2015}, supporting the paradigm in which stars form within gravitationally unstable filaments \citep{Andre+2014}. The core mass function (CMF) closely resembles the IMF \citep{Motte+1998, Konyves+2015, Konyves+2020, Li+2023}, suggesting the slope is set already at core formation. If the CMF arises from filament fragmentation \citep{Andre+2019}, the IMF slope should be linked to that of the filament linear density function (FLDF) -- a hypothesis that requires measuring the FLDF across a large, diverse cloud sample.

A key quantity governing fragmentation is the linear density $\Lambda$ (mass per unit length): filaments exceeding the critical value $\Lambda_{\rm c}$ (supercritical) are gravitationally unstable and fragment into dense cores \citep{Ostriker1964, Inutsuka+Miyama1992}, a criterion applied to observed filaments \citep[e.g.,][]{Zhang+2020}. However, $\Lambda$ is not the sole controlling parameter; turbulence, magnetic fields, geometry, and environmental conditions also influence fragmentation \citep{Mathew+Federrath2021}. Filaments are hierarchically substructured, with larger-scale filaments containing intertwined smaller-scale sub-filaments or fibers \citep{Hacar+2013, Hacar+2018}, so $\Lambda$ and the FLDF are inherently scale-dependent. The FLDF has been measured in nearby clouds \citep{Andre+2019, Zhang+2024}, but not yet systematically with independent multiscale extraction.

\begin{figure}
\centering
\includegraphics[width=1.0\textwidth]{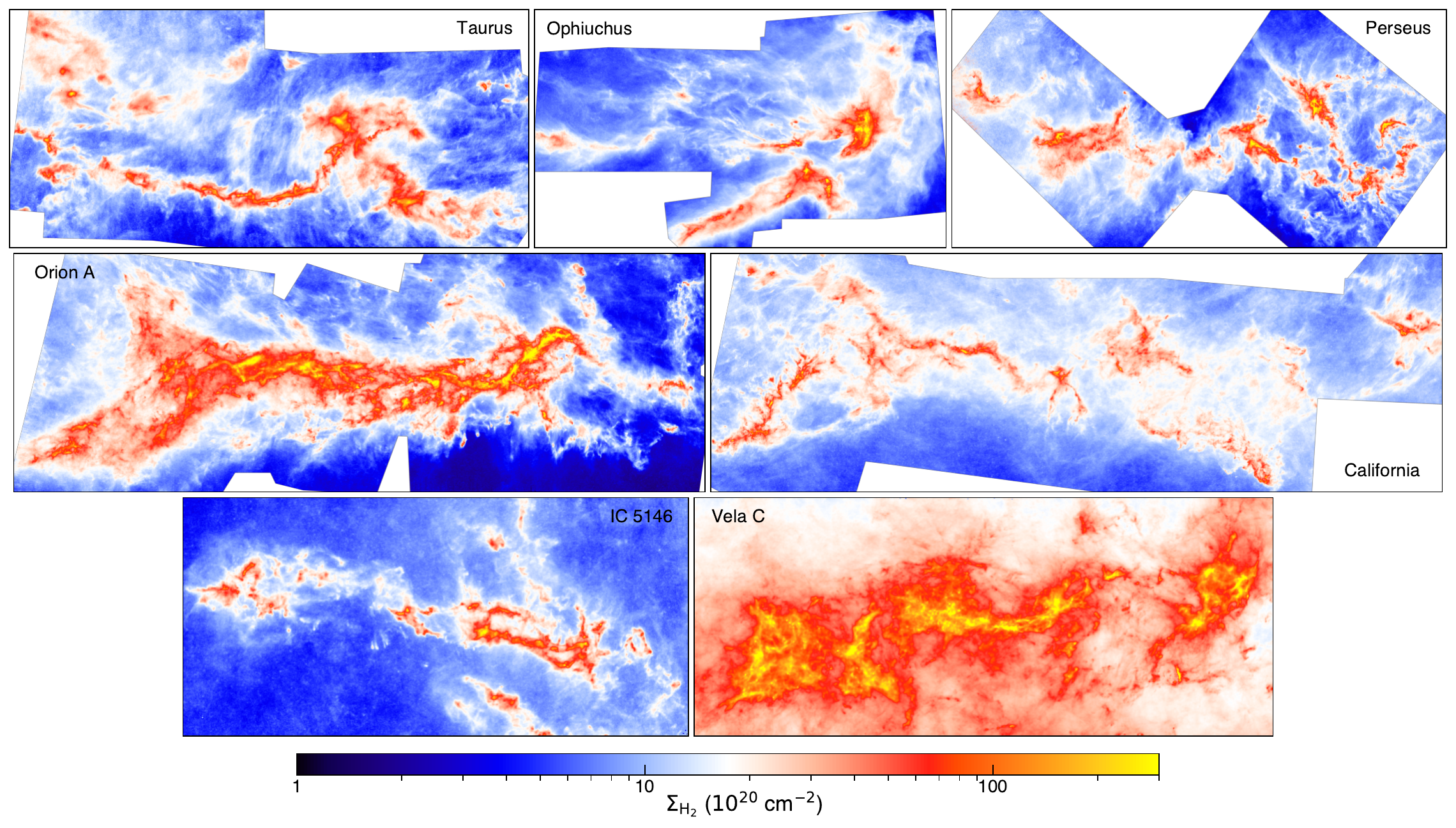}
\caption{Shown are high-resolution ($13.5^{\prime\prime}$) surface density maps derived from \textit{Herschel} PACS and SPIRE images observed in several surveys: HGBS \citep{Andre+2010} for Taurus, Ophiuchus, Perseus, Orion\,A, and IC\,5146; HOBYS \citep{Motte+2010} for Vela\,C; and A-CMC \citep{Harvey+2013} for California.
The horizontal color bar shows the surface density on a logarithmic scale, in units of $10^{20}$\,cm$^{-2}$, as indicated.
}
\label{fig:overview}
\end{figure}

\begin{figure}
\centering
\includegraphics[width=1.0\textwidth]{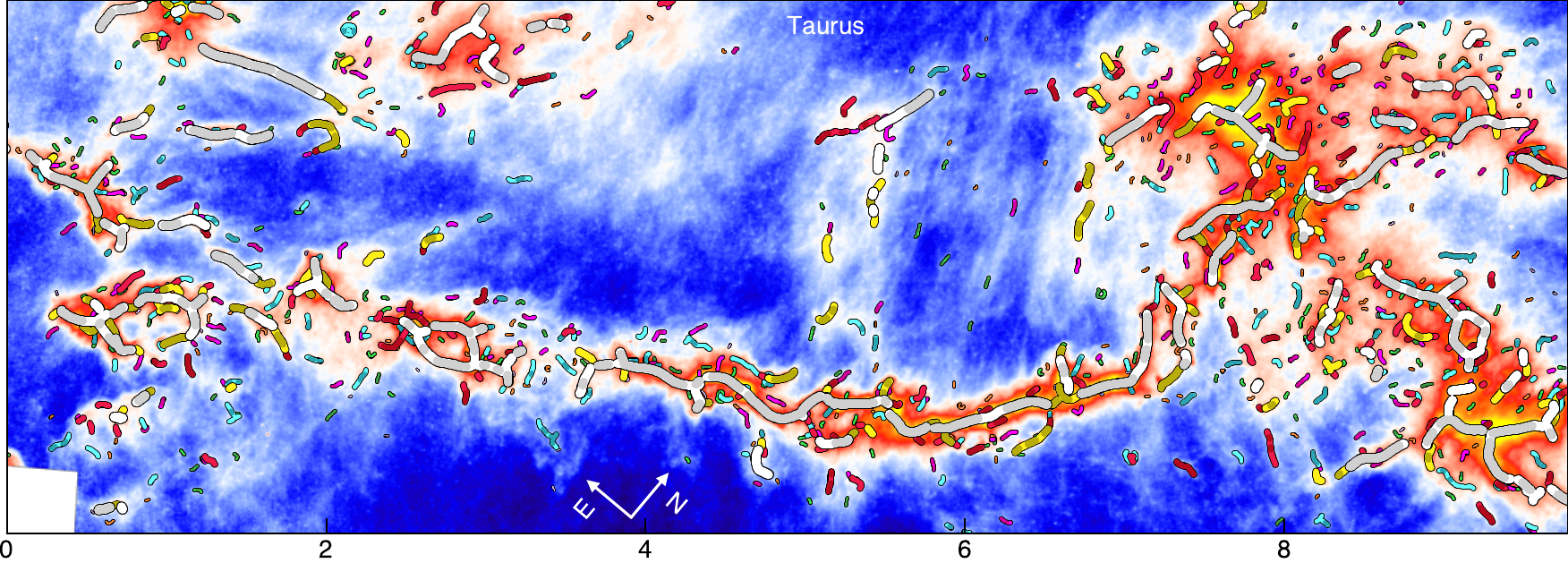}
\includegraphics[width=1.0\textwidth]{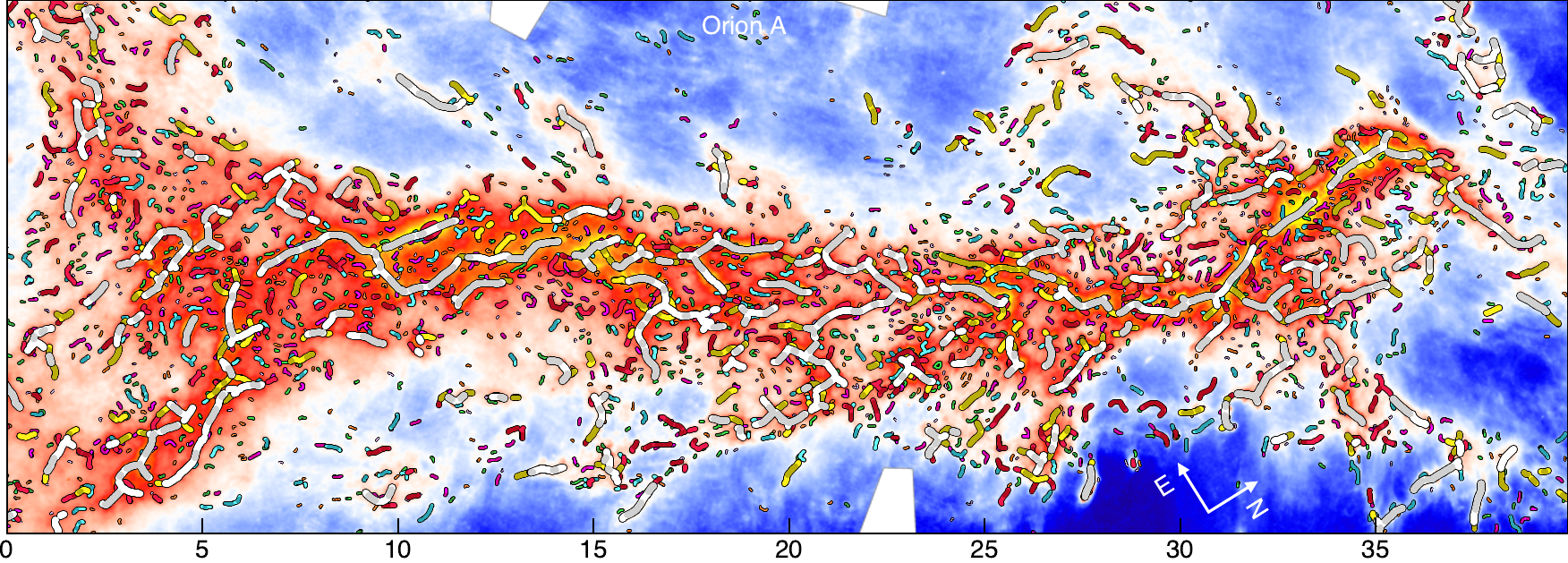}
\includegraphics[width=1.0\textwidth]{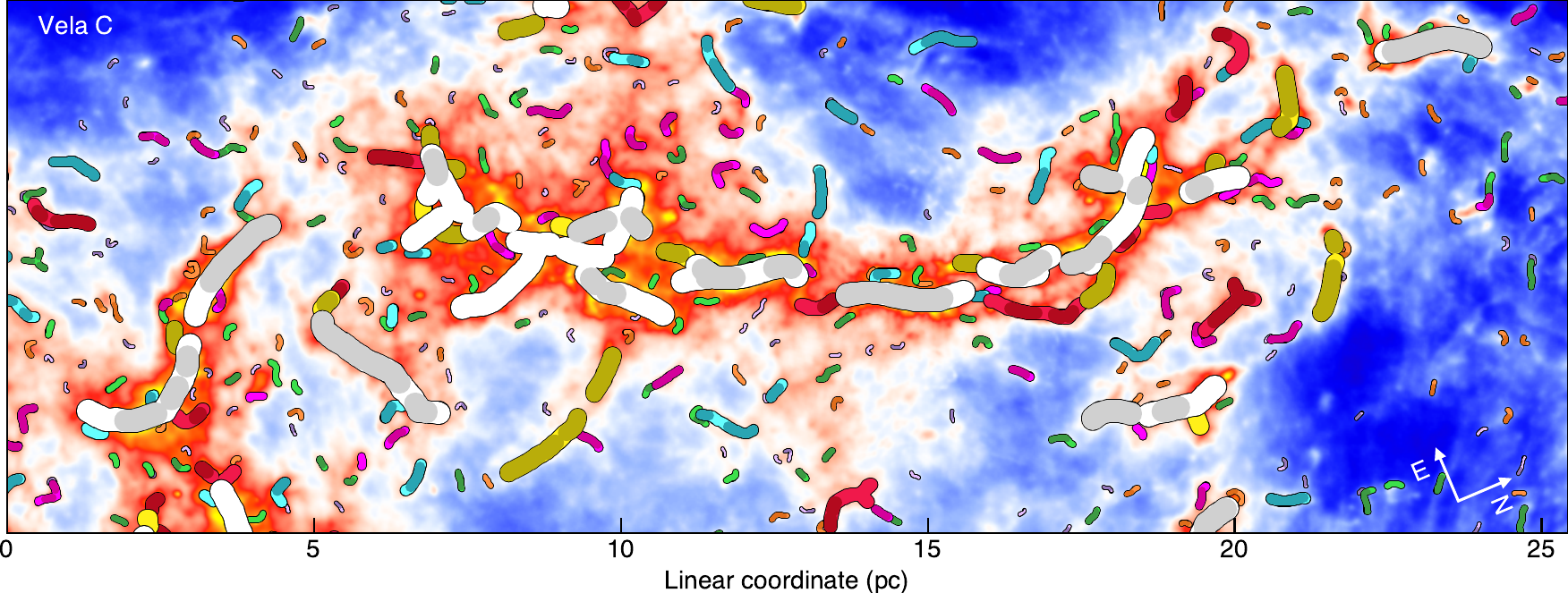}
\includegraphics[width=1.0\textwidth]{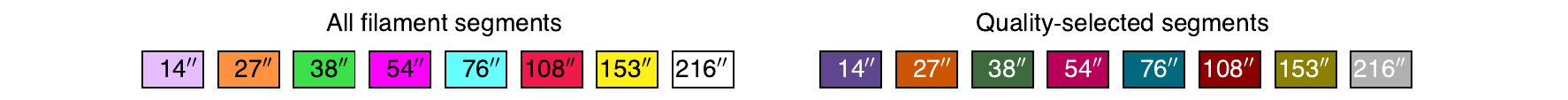}
\caption{Surface density images of three nearby molecular clouds -- Taurus, Orion~A, and Vela~C -- overlaid with scale-dependent filament skeletons extracted by \textit{getsf} on spatial scales from 14 to 216$^{\prime\prime}$. Skeleton line widths are proportional to the detection scale, as indicated in the colorbar. In each panel, lighter hues represent all detected filament segments, while the darker shades of the corresponding hues mark those segments that satisfy the profile quality criteria. The combined colorbar at the bottom explicitly distinguishes the two categories.
}
\label{fig:filament_skeletons}
\end{figure}

Here we present the first FLDF measurement across seven nearby clouds -- Taurus, Ophiuchus, Perseus, Orion~A, California, IC~5146, and Vela~C (Figure~\ref{fig:overview}) -- 
spanning cloud distances of 140--920\,pc from the Sun and a wide range of activity, using \textit{getsf} \citep{Menshchikov2021method}, which traces filament skeletons independently on eight scales from 14 to 216$^{\prime\prime}$ (Figure~\ref{fig:filament_skeletons}), with the improved profile fitting of \citet{Menshchikov+Zhang2026subm}. Combining all clouds and scales yields a statistically robust composite FLDF whose slope we compare directly with Salpeter's. A companion paper \citep{Zhang+2026} analyzes filament widths and volume density profiles for the same sample; here we focus on the linear-density distribution and its link to the IMF.

\section{Scale-dependent linear density distributions}

\begin{figure}
\centering
\includegraphics[width=1.0\textwidth]{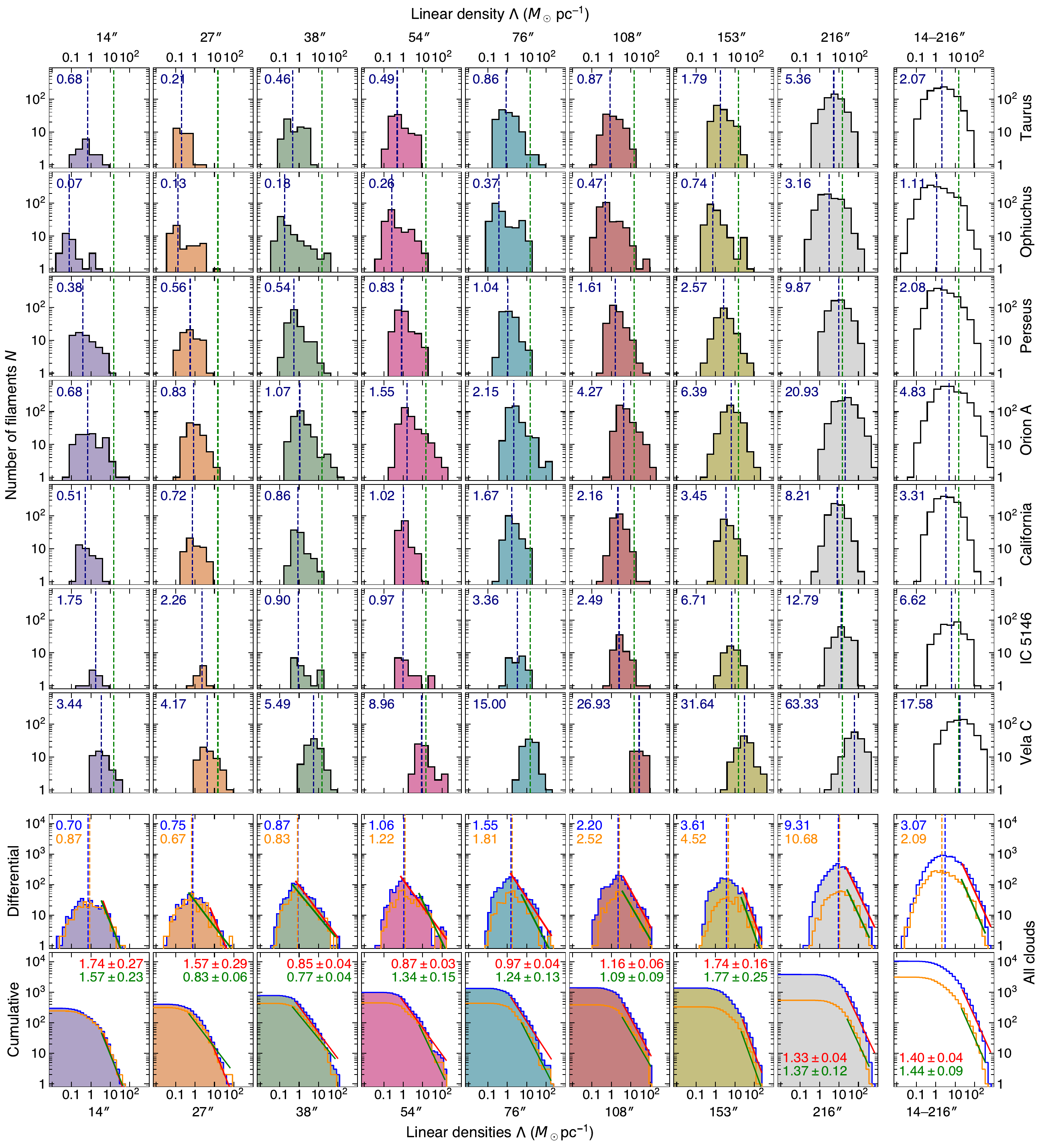}
\caption{Top seven rows: per-scale linear-density distributions for each cloud; dashed lines mark the medians $\tilde{\Lambda}_{k}$ (black) and the critical $\Lambda_{\rm c}=15\,M_\odot$\,pc$^{-1}$ (green). Bottom rows: differential ($N$) and cumulative ($N_{\rm C}$) distributions combined over all clouds, for filament segments (scale-colored fills, using the same deep hues as the quality‑selected segments in Figure~\ref{fig:filament_skeletons}, with blue dashed medians) and entire filaments (orange). Red and green lines are maximum-likelihood (MLE) power-law fits to the segment and entire‑filament distributions above the fitted lower bound $\Lambda_{\rm min}$.
}
\label{fig:lindensDistribution}
\end{figure}

Figure~\ref{fig:lindensDistribution} shows the scale-dependent linear-density distributions $\Lambda_{k}$ derived from the surface density profiles (Equation~\ref{surfdens_fittingfun}), together with the all-scales accumulated distributions. The bottom two rows give the differential and cumulative distributions combined over all clouds, with the entire-filament (crest-averaged) distributions shown for comparison.

\begin{figure}
\centering
\includegraphics[width=0.45\textwidth]{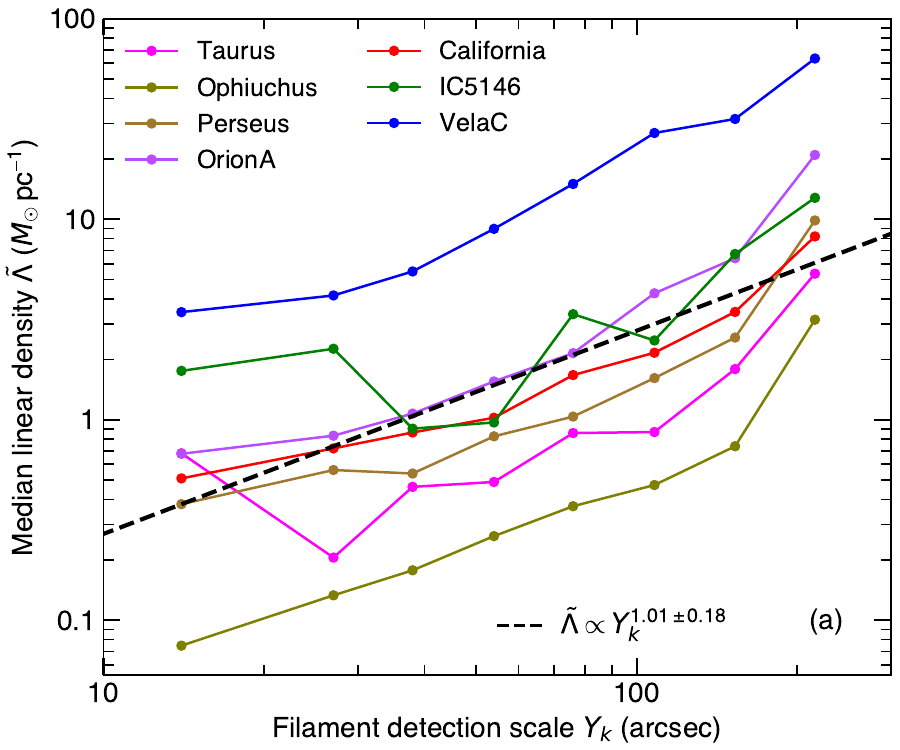}
\includegraphics[width=0.45\textwidth]{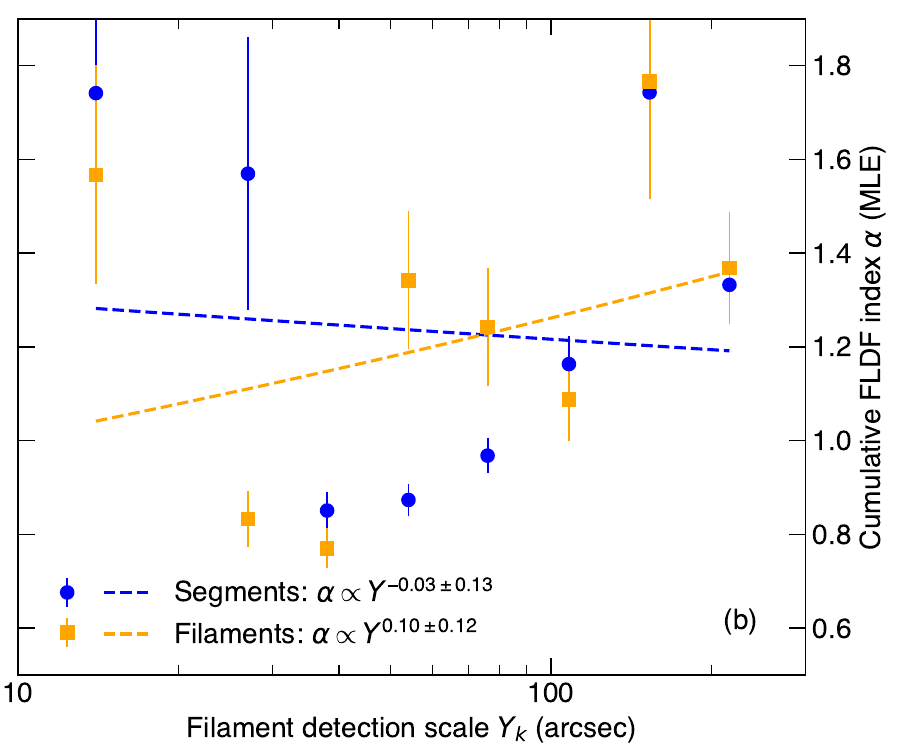}
\caption{(a) Median linear density $\tilde{\Lambda}$ versus detection scale $Y_k$ for the seven clouds (colored lines); the dashed line is the combined fit $\tilde{\Lambda} \propto Y_k^{1.01 \pm 0.18}$. The cloud-to-cloud spread reflects genuine differences in network mass content, and the steeper rise on the two largest scales (153--216$^{\prime\prime}$) may indicate a growing envelope contribution. (b) FLDF index $\alpha$ from the high-$\Lambda$ end of the cumulative distributions (Figure~\ref{fig:lindensDistribution}): blue circles, filament segments; orange squares, whole filaments; error bars, MLE standard error $\alpha/\sqrt{n}$, where $n$ is the number of points in the fitted tail (Appendix~\ref{plawfittingFLDF})}. Dashed lines are power-law fits with the best-fit exponents indicated.
\label{fig:scale_relations}
\end{figure}

Filaments detected on larger scales have systematically higher linear densities: across the seven clouds, the median $\tilde{\Lambda}_{k}$ spans 0.07--3.4\,$M_{\odot}$\,pc$^{-1}$ on the 14$^{\prime\prime}$ scale and 3.2--63\,$M_{\odot}$\,pc$^{-1}$ on the 216$^{\prime\prime}$ scale, following $\tilde{\Lambda} \propto Y_k^{1.01 \pm 0.18}$ (Figure~\ref{fig:scale_relations}a; per-cloud exponents 0.72--1.2). Combined over all clouds, the median shifts from 0.70 to 9.3\,$M_{\odot}$\,pc$^{-1}$ between the same scales ($\tilde{\Lambda} \propto Y^{0.91 \pm 0.16}$). Vela~C shows the highest medians on all scales and Ophiuchus the lowest, reflecting genuine differences in their networks' mass content.

We constructed the composite FLDF by combining linear densities of all filament segments from all seven clouds and eight spatial scales. Because \textit{getsf} extracts filaments independently on each scale, a large-scale filament and the smaller-scale sub-filaments nested within it are treated as separate structures on their respective scales. We estimate every FLDF slope by unbinned maximum-likelihood fitting above a lower bound $\Lambda_{\rm min}$ (Appendix~D). The cumulative FLDF slope shows no significant trend with scale ($\alpha \propto Y^{0.03 \pm 0.13}$; Figure~\ref{fig:scale_relations}b).

\section{A Salpeter-like composite FLDF}

The composite FLDF, combining all clouds and scales, follows ${\rm d}N/{\rm d}\log\Lambda \propto \Lambda^{-\alpha}$ with $\alpha = 1.40 \pm 0.04$ (filament segments) and $1.44 \pm 0.09$ (entire filaments), close to the Salpeter slope of $1.35$ \citep{Salpeter1955}. The per-scale slopes show no significant trend with scale, scattering about this composite value (Figure~\ref{fig:scale_relations}b); the Salpeter-like slope is therefore an approximately scale-independent property of the filamentary mass distribution. The composite follows a single power law over more than a decade in $\Lambda$, with the combined median linear density rising coherently with scale ($\tilde{\Lambda} \propto Y^{0.9}$) while the slope stays consistent. Unlike the single $\sim$0.1\,pc width emphasized previously \citep{Arzoumanian+2011, Arzoumanian+2019}, our multiscale analysis reveals a hierarchy of widths and linear densities that any model of the IMF slope must accommodate. Because the composite is fit above $\Lambda_{\rm min} \approx 23\,M_\odot$\,pc$^{-1}$, well above $\Lambda_{\rm c}$, the Salpeter-like slope is determined by the supercritical filament backbones -- the structures most directly relevant to star formation -- rather than by the more numerous subcritical sub-structures.

Our result agrees with the single California cloud measurement of \citet{Zhang+2024} using the same \textit{getsf} method and extends it to a diverse sample. It is shallower than the $\alpha \approx 1.6 \pm 0.1$ of \citet{Andre+2019}, who used \textit{disperse} skeletons and crest-averaged profiles. The difference likely reflects these methodological choices, as \textit{getsf} traces scale-dependent skeletons and fits individual filament segments (Equation C1; the two methods are compared in Appendix A of \citealt{Zhang+2026}).

The Salpeter-like slope distinguishes filaments from more diffuse molecular cloud structures, whose mass function is significantly shallower (${\rm d}N/{\rm d}\log M \propto M^{-0.7}$, \citealp{Blitz1993, Kramer+1998}). This contrast is suggestive of a distinct evolutionary step: starting from the shallow $M^{-1}$ distribution expected from scale-free turbulent fragmentation \citep{Elmegreen1997}, gravitational accretion onto supercritical filaments ($\dot{\Lambda} \propto \sqrt{\Lambda}$, \citealp{Andre+2019}) naturally steepens the FLDF to the observed Salpeter value \citep{Hennebelle+Andre2013} on a timescale of $\sim$0.5--1\,Myr \citep{Palmeirim+2013, Andre+2019}.

Although star formation activity varies strongly across the sample -- supercritical-filament fractions range from $\sim$7\% in the quiescent clouds (Taurus, Ophiuchus) to $\sim$54\% in Vela~C, tracking independent activity indicators \citep{Evans+2009, Lada+2010, Zhang+2026} -- the composite slope stays near Salpeter's value in every cloud, making it an environment-independent property of the integrated filamentary mass distribution. In Vela~C, supercritical filaments host exclusively prestellar and protostellar cores \citep{Li+2023}, confirming that the largest, most massive structures dominate the high-$\Lambda$ tail and hence the composite slope.

\begin{figure}
\centering
\includegraphics[width=0.9\textwidth]{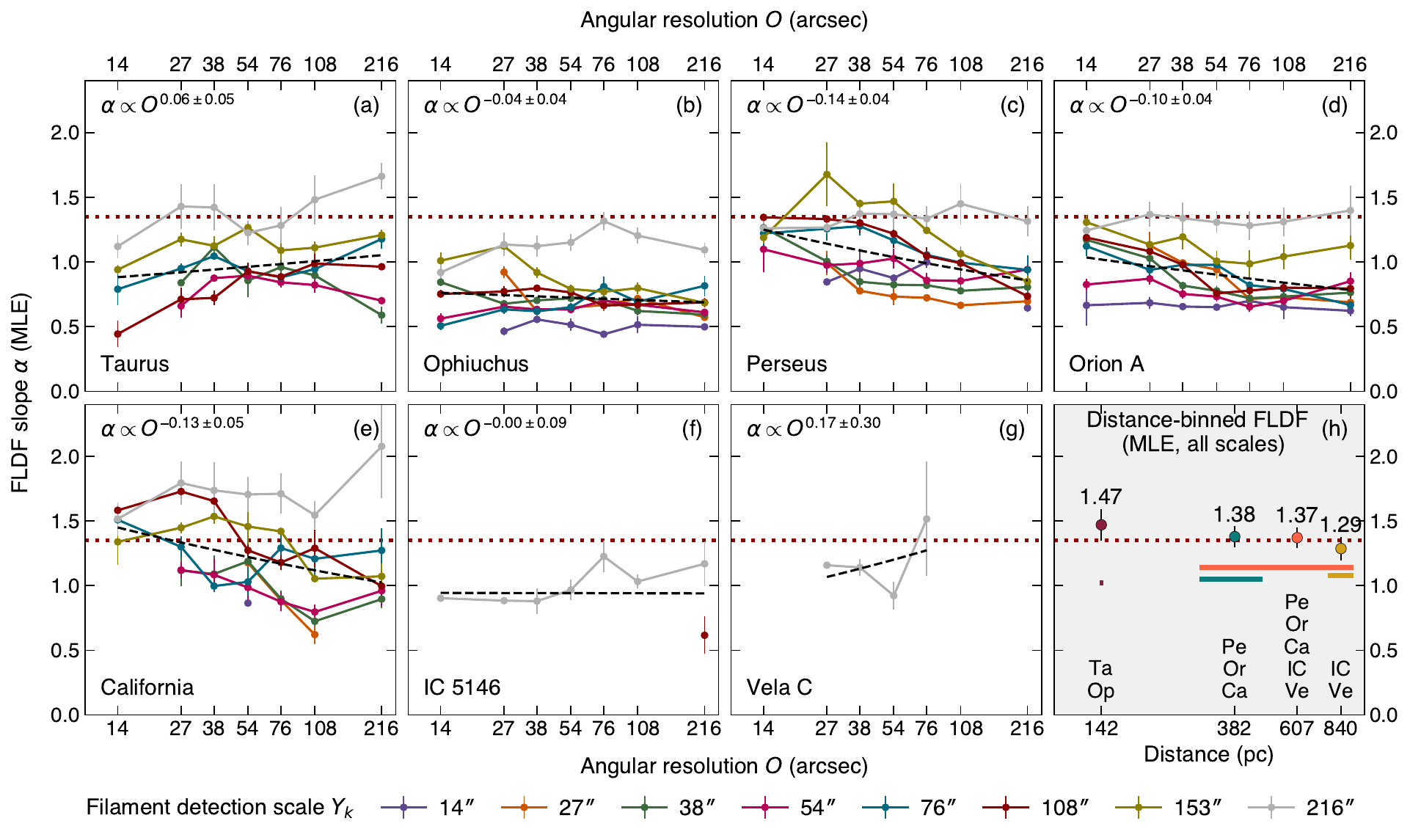}
\caption{(a)--(g) FLDF index $\alpha$ versus angular resolution $O$ for each cloud; colors denote detection scales $Y_k$ (14--216$^{\prime\prime}$). Dotted line: Salpeter slope $\alpha = 1.35$ \citep{Salpeter1955}. The absence of steepening with $O$ shows that blending raises measured linear densities but not the slope \citep{Zhang+2026}. (h) Composite slopes for distance-binned sub-samples (Ta+Op, Pe+Or+Ca, IC+Ve, and Pe+Or+Ca+IC+Ve), plotted at median distances with bars spanning the distance ranges; all are consistent with Salpeter's value, independent of distance.}
\label{fig:FLDF_robustness}
\end{figure}

The Salpeter-like slope is robust against observational biases. Because all \textit{Herschel} maps share a $13.5^{\prime\prime}$ resolution, more distant clouds are probed at coarser physical resolution, blending structures and raising measured linear densities \citep{Zhang+2026}. Convolving each map to coarser resolutions $O$ from 14 to 216$^{\prime\prime}$ and re-measuring $\alpha$ on every detection scale $Y_k$ (Figure~\ref{fig:FLDF_robustness}a--g) shows no systematic steepening with $O$: blending raises linear densities but not the slope. Distance-binned composite slopes (Figure~\ref{fig:FLDF_robustness}h) range from $1.47 \pm 0.12$ (Taurus\,+\,Ophiuchus, 140--144\,pc) to $1.29 \pm 0.09$ (IC~5146\,+\,Vela~C, 760--920\,pc) and stay mutually consistent across a factor of $\sim$7 in distance; the slight decrease with distance is opposite to blending-driven steepening, further excluding an artifact.

Our conclusions rest on quantities that are largely independent of the extraction method. Many filament-identification schemes exist, from intensity- and template-based tracers to topological, persistence-based methods, and they need not return identical slopes: the \textit{disperse}-based measurement of \citet{Andre+2019} gives a steeper $\alpha \approx 1.6$, a difference traceable to crest-averaging rather than our scale-dependent segment fitting (Appendix~A of \citealt{Zhang+2026}). Both approaches nonetheless yield steep, Salpeter-like slopes, and our result is anchored by features that do not depend on the algorithm: the median linear density rises with scale as $\tilde{\Lambda} \propto Y$ in every cloud, and the composite slope is stable against changes in angular resolution and cloud distance (Figure~\ref{fig:FLDF_robustness}). A systematic comparison across extraction methods is beyond the scope of this Letter, but these checks indicate that the Salpeter-like composite FLDF reflects the filamentary mass distribution rather than the \textit{getsf} algorithm.

\section{Linking the FLDF to the core mass function}

The FLDF can be quantitatively linked to the CMF (${\rm d}N/{\rm d}\log M \propto M^{-\psi}$) through the fragmentation process. A central tenet of the filament fragmentation paradigm is that dense cores form via gravitational fragmentation of filaments \citep{Inutsuka+Miyama1992, Andre+2014}. Consider a filament segment of length $\ell$ and linear density $\Lambda$: if fragmentation occurs on a characteristic length $\lambda_{\rm fr}$, the number of cores produced is $\ell/\lambda_{\rm fr}$ and the typical core mass is $M \sim \Lambda\,\lambda_{\rm fr}$. Thus the CMF is a transformed version of the FLDF, with the transformation determined by how $\lambda_{\rm fr}$ depends on $\Lambda$. Assuming the minimal power-law parameterization $\lambda_{\rm fr} \propto \Lambda^{\eta}$, the core mass varies as $M \propto \Lambda^{1+\eta}$. Since the FLDF gives ${\rm d}N/{\rm d}\Lambda \propto \Lambda^{-\alpha-1}$, a standard change of variables yields the CMF slope:
\begin{equation}
\psi = \frac{\alpha + \eta}{1 + \eta}.
\label{eq:psi_cmf}
\end{equation}

This simple power‑law parameterization is intended as an illustrative framework; in reality, the fragmentation length may depend on additional factors such as turbulence, magnetic fields, and the detailed density profile, and the mapping between FLDF and CMF could be more complex than Equation~(\ref{eq:psi_cmf}) suggests. The exponent $\eta$ is not free but encodes the fragmentation regime. For an isothermal self-gravitating cylinder, perturbation theory gives the most unstable wavelength $\lambda_{\rm fr} \propto c_{\rm s}^2/(G\Lambda)$ \citep{Inutsuka+Miyama1992}, i.e.\ $\eta = -1$: denser filaments fragment on shorter lengths. If instead the fragmentation length follows the filament width, our measured $\tilde{H} \propto \tilde{\Lambda}^{0.5}$ \citep{Zhang+2026} implies $\eta = 0.5$, although the link between $\lambda_{\rm fr}$ and $H$ is uncertain and this is only a consistency check. For $\eta = 0$ the fragmentation length is independent of $\Lambda$ -- set, e.g., by the sonic or turbulence correlation length \citep{Padoan+Nordlund2002, Federrath2016} -- and the CMF directly inherits the FLDF slope, $\psi = \alpha$; this requires only that the most unstable wavelength decouple from $\Lambda$ (as when turbulence rather than gravity sets it), not that widths be constant. The ansatz $\lambda_{\rm fr} \propto \Lambda^{\eta}$ thus spans pure gravity ($\eta=-1$), fixed length ($\eta=0$), and intermediate cases.

The case $\eta = 0$ is consistent with the CMF in nearby clouds ($\psi \approx 1.3$--$1.4$, \citealp{Konyves+2015, Konyves+2020}), providing the quantitative link between the Salpeter-like FLDF and the Salpeter-like CMF; we note however that this comparison is indirect, as the FLDF and CMF are measured from different physical structures -- filament segments and dense cores respectively -- and a direct test requires measuring both distributions from the same clouds. We also note that the derivation assumes a roughly constant fragmentation efficiency per unit filament length; if efficiency varies with $\Lambda$, an additional factor would modify Equation~(\ref{eq:psi_cmf}). The latter thus provides a unified framework: the same FLDF can produce different CMF slopes depending on the dynamical environment, with $\eta$ as the key empirical parameter connecting them.

\section{The high-mass protocluster tension}

Several caveats apply to the FLDF--CMF--IMF chain: the fragmentation length, core formation efficiency, and the roles of magnetic fields and turbulence remain uncertain \citep{Federrath2016, Hennebelle2013}, and the CMF to IMF mapping is not one-to-one, with protostellar feedback \citep{Adams+Fatuzzo1996, Dib+2011} and competitive accretion \citep{Bonnell+2001, Kroupa2008} reshaping the high-mass end. Direct, simultaneous FLDF and CMF measurements in the same clouds are needed.

Recent ALMA observations of high-mass protoclusters report CMF slopes well below Salpeter value: the ALMA-IMF program finds a combined ${\rm d}N/{\rm d}\log M \propto M^{-0.97}$ across 15 protoclusters \citep{Louvet+2024}. Yet Equation~(\ref{eq:psi_cmf}) gives $\psi - 1 = (\alpha - 1)/(1 + \eta)$, so for $\alpha > 1$, $\psi$ approaches unity from above as $\eta \rightarrow \infty$ and never falls below it for any $\eta > -1$. With $\alpha \approx 1.40$ and $\eta = 0.5$ (from $\tilde{H} \propto \tilde{\Lambda}^{0.5}$ with $\lambda_{\rm fr} \propto H$; \citealt{Zhang+2026}) the predicted $\psi \approx 1.27$ -- still steeper than Salpeter slope and incompatible with the ALMA-IMF value.

The shallow ALMA-IMF CMF therefore cannot be reconciled with our Salpeter-like FLDF by width-regulated fragmentation alone. Possible resolutions are (i) a genuinely different, $\alpha < 1$ FLDF in massive protoclusters; (ii) a fragmentation length not following a single power law in $\Lambda$; (iii) a $\Lambda$-dependent fragmentation efficiency, adding a multiplicative factor to Equation~(\ref{eq:psi_cmf}); or (iv) systematic core-mass uncertainties from assumed dust temperatures and background subtraction. A natural possibility is that continued subfragmentation steepens the mass function toward Salpeter slope; however, a hierarchical analysis of W43-MM1 from 14\,kau to 270\,au shows it stays top-heavy ($\psi \approx 0.92$) even on disk scales \citep{Motte+2026subm}, ruling this out. The tension thus remains unresolved, and discriminating among (i)--(iv) requires simultaneous FLDF and CMF measurements in the same protoclusters.

\section{A unified picture for the Salpeter slope}

We emphasize that the similarity between the FLDF slope and the Salpeter
value, while suggestive, does not by itself establish a direct causal link
to the IMF.  The interpretation that the FLDF slope explains the origin of
the Salpeter IMF rests on additional assumptions concerning the fragmentation
process, the core-to-star mapping, and the role of feedback.  The FLDF
measurement is a robust observational result; the proposed connection to the
IMF remains a physically motivated but testable hypothesis.
The observational result of a Salpeter-like FLDF suggests a possible unified picture of the IMF, in which: (i) turbulence generates filaments with a shallow, scale-free mass distribution; (ii) gravitational accretion steepens the FLDF toward Salpeter slope; (iii) fragmentation with $\eta \approx 0$ yields a CMF inheriting this slope \citep{Andre+2019}; and (iv) core collapse produces the final IMF \citep{Konyves+2015}. While alternative interpretations cannot be excluded by the FLDF measurement alone, the close agreement between the composite slope ($\alpha \approx 1.40$) and the Salpeter value is striking. If the unified picture outlined above is correct, the composite slope emerges from integrating a hierarchical filament population whose median linear density grows linearly with scale. In this scenario, the Salpeter slope would originate in the scale-invariant filamentary mass distribution combined with a fragmentation length independent of linear density, making the IMF's universality a consequence of the global, multiscale process of filament assembly and collapse rather than of local conditions. Equation~(\ref{eq:psi_cmf}) offers a quantitative test: $\eta$ can be measured wherever both the FLDF and CMF are available, probing directly how the fragmentation length depends on linear density.

\begin{acknowledgments}
This work is based on observations obtained with \textit{Herschel}, an ESA space observatory with science instruments provided by European-led Principal Investigator consortia and with important participation from NASA. Images of the molecular clouds were obtained in the \textit{Herschel} Gould Belt Survey (HGBS, PI: Ph.~Andr{\'e}), the HOBYS program (PI: F.~Motte, A.~Zavagno, S.~Bontemps), and the A-CMC survey (PI: P.~Harvey). HGBS and HOBYS are \textit{Herschel} Key Projects jointly carried out by SPIRE Specialist Astronomy Group 3 (SAG3), scientists of several institutes in the PACS Consortium, and scientists of the \textit{Herschel} Science Center (HSC). This work was supported by the Key Project of International Cooperation of the Ministry of Science and Technology of China through grant 2010DFA02710, and by the National Natural Science Foundation of China through grants 11503035, 11573036, 11373009, 11433008, 11403040, and 11403041. G.-Y. Zhang acknowledges support from the Postdoctoral Science Foundation of China (No. 2021T140672).

The \textit{Herschel} data are publicly available from the \textit{Herschel} Science Archive (\url{http://archives.esac.esa.int/hsa/whsa/}). Derived surface density maps and filament catalogs are available from the corresponding author upon request. The \textit{getsf} filament extraction software is available at \url{http://irfu.cea.fr/Pisp/alexander.menshchikov/}. The Python code implementing the profile fitting function (Equation~\ref{surfdens_fittingfun}), the analytical relations of \citet{Menshchikov+Zhang2026subm}, and the fitting strategy used to derive filament linear densities is deposited on Zenodo~\citep{Zhang+Menshchikov2026code} (\url{https://doi.org/10.5281/zenodo.20293702}).
\end{acknowledgments}

\facilities{\textit{Herschel} (PACS, SPIRE)}

\software{\textit{getsf} \citep{Menshchikov2021method}, \textit{swarp} \citep{Bertin+2002}}

\appendix

\section{Observational data and surface density maps}

We used archival \textit{Herschel} data\footnote{\url{http://archives.esac.esa.int/hsa/whsa/}} for Taurus, Ophiuchus, Perseus, Orion~A, California, IC~5146, and Vela~C, comprising 70--500\,$\mu$m imaging (resolutions 8.4--36.3$^{\prime\prime}$) from the Gould Belt and HOBYS surveys \citep{Andre+2010, Motte+2010, Harvey+2013}, resampled to a common 3$^{\prime\prime}$ pixel scale with \textit{swarp} \citep{Bertin+2002}. Adopted distances are 140, 144, 294, 432, 470, 760, and 920\,pc, respectively \citep{Zucker+2019, Zucker+2020}.

We derived H$_2$ surface density maps with the \textit{hires} algorithm \citep{Menshchikov2021method}, fitting each pixel's spectral energy distribution with a modified blackbody (optically thin emission, $\kappa_\nu \propto \nu^2$; \citealt{Hildebrand1983, Ossenkopf+Henning1994}; dust-to-gas ratio 0.01). Zero-level offsets were set by comparison with \textit{Planck} \citep{Bernard+2010, PlanckCollaboration+2014}. The maps have an effective resolution of 13.5$^{\prime\prime}$.

\section{Multiscale filament extraction}

Filaments were extracted with \textit{getsf}\footnote{\url{http://irfu.cea.fr/Pisp/alexander.menshchikov/}} \citep{Menshchikov2021method} on eight scales $Y_k =$ 14, 27, 38, 54, 76, 108, 153, and 216$^{\prime\prime}$ (a subscript $k$ denotes a quantity measured on scale $Y_k$). The method decomposes images into single-scale components, removes compact sources, and traces filament skeletons independently on each scale, so that structures much smaller or larger than $Y_k$ do not affect them; the skeletons represent filaments whose half-maximum widths match $Y_k$ (Figure~\ref{fig:filament_skeletons}). Skeletons were divided into 10-pixel (30$^{\prime\prime}$) segments to capture variations along their lengths.

\section{Deriving linear densities from filament profiles}

We applied the profile fitting method developed by \citet{Menshchikov+Zhang2026subm}, which accounts for the finite outer radius of filamentary structures. The surface density profile fitting function is given by
\begin{equation}
\Sigma(r) = \Sigma_{\rm C}\left(1+\left(2^{2/\gamma}-1\right)\left(\frac{2r}{w}\right)^2\right)^{-\gamma/2}\left(1-\left(\frac{r}{R}\right)^{\epsilon}\right)^{1/2},
\label{surfdens_fittingfun}
\end{equation}
where $\Sigma_{\rm C}$ is the crest surface density, $\gamma$ the intrinsic slope, $w$ the intrinsic half-maximum width, $R$ the boundary radius, and $\epsilon$ an empirical exponent enforcing consistency with the volume density profile. The fit optimizes three free parameters ($\gamma$, $R$, $\Sigma_{\rm C}$); $w$ and $\epsilon$ are fixed by $\gamma$, $R$, and the measured width $H$ through self-consistency relations between $\Sigma(r)$ and $\rho(r)$, given explicitly in \citet{Menshchikov+Zhang2026subm} and implemented in the public fitting code \citep{Zhang+Menshchikov2026code}. Residual backgrounds were subtracted by linear interpolation between profile endpoints. We retained only profiles with $R_{\rm d}^2 > 0.97$, relative residuals below 0.2, and well-constrained $\gamma$ (excluding those at the fitting bound $\gamma = 8$).

The linear densities were computed by numerical integration of the fitted surface density profiles:
\begin{equation}
\Lambda_{k} = 2\mu m_{\rm H}\int_{0}^{R}\Sigma_{k}(r)\,{\rm d}r,
\label{fil_linear_densities}
\end{equation}
where $\mu = 2.8$ is the mean molecular weight per H$_2$ molecule and $m_{\rm H}$ is the mass of the hydrogen atom. Linear densities derived independently from volume density profiles agree with the surface density estimates within a median relative difference of 1\% \citep{Zhang+2026}, confirming the robustness of the results presented here.

\section{Power-law fitting of the FLDF}
\label{plawfittingFLDF}

We estimated all FLDF slopes by unbinned maximum likelihood \citep{Clauset+2009}, working directly with the individual linear densities $\Lambda_i$ above a lower bound $\Lambda_{\rm min}$. For the continuous power law defined through ${\rm d}N/{\rm d}\log\Lambda \propto \Lambda^{-\alpha}$, the maximum-likelihood slope is $\hat{\alpha} = n / \sum_i \ln(\Lambda_i/\Lambda_{\rm min})$, with standard error $\sigma = \hat{\alpha}/\sqrt{n}$, where $n$ is the number of segments above $\Lambda_{\rm min}$. We selected $\Lambda_{\rm min}$ for each fit by minimizing the Kolmogorov--Smirnov distance between the empirical and fitted cumulative distributions \citep{Clauset+2009}; 
for the composite distribution this yields 
$\Lambda_{\rm min} \approx 23\,M_\odot$\,pc$^{-1}$, above the critical 
value $\Lambda_{\rm c} \approx 15\,M_\odot$\,pc$^{-1}$, so the composite 
slope is set by supercritical filaments.
The slopes are insensitive to reasonable variations in $\Lambda_{\rm min}$. Unbinned maximum likelihood avoids the bias and inflated uncertainties that affect least-squares fits to binned or cumulative counts.

\bibliographystyle{aasjournalv7}
\bibliography{ApJL_ref}

\end{document}